\PassOptionsToPackage{square, comma, sort&compress}{natbib}
\documentclass[preprint,12pt]{elsarticle}

\usepackage{amsmath}
\usepackage[colorlinks=true,linkcolor=blue,urlcolor=blue,citecolor=blue]{hyperref}
\usepackage{graphicx}
\usepackage[squaren, Gray, cdot]{SIunits}
\usepackage{amssymb}

\usepackage{tikz}
\begin{document}

\begin{frontmatter}

\title{Observation of magnetically-induced transition intensity redistribution in the onset of the hyperfine Paschen-Back regime
%Symmetry breaking of circularly-polarized light absorption spectrum in the Hyperfine Paschen-Back regime
% Effect of polarization on strong changes of Cs atomic transition intensities within the groups $F_g=3 \rightarrow F_e=5$ and $F_g=4 \rightarrow F_e=2$ in strong magnetic fields 
}
% %\tnotetext[mytitlenote]{Fully documented templates are available in the elsarticle package on \href{http://www.ctan.org/tex-archive/macros/latex/contrib/elsarticle}{CTAN}.}

\author[IPR]{Armen Sargsyan}
% \author[IPR]{Arevik Amiryan}
\author[UFC]{Emmanuel Klinger\corref{mycorrespondingauthor}}
\cortext[mycorrespondingauthor]{emmanuel.klinger@femto-st.fr}
% \ead{}
\author[IPR]{Ara Tonoyan}

\author[IPR]{David Sarkisyan}

\address[IPR]{Institute for Physical Research – National Academy of Sciences of Armenia, 0203 Ashtarak-2, Armenia}
\address[UFC]{Universit\'e de Franche-Comt\'e, SupMicroTech-ENSMM, UMR 6174 CNRS, Institut FEMTO-ST, 25030 Besan\c{c}on, France}

\begin{abstract}

The Zeeman effect is an important topic in atomic spectroscopy. The induced change in transition frequencies and amplitudes finds applications in the Earth-field-range magnetometry.
%Depending on the magnitude of the magnetic field, the effect is referred to as non-linear Zeeman shift (low field) or linear Zeeman (high field).
At intermediate magnetic field amplitude $B\sim B_0 = A_\text{hfs}/\mu_B$, where $A_\text{hfs}$ is the magnetic dipole constant of the ground state, and $\mu_B$ is the Bohr magneton ($B_0\approx 1.7$\,kG for Cs), the rigorous rule $\Delta F = 0, \pm1$ is affected by the coupling between magnetic sub-levels induced by the field. 
Transitions satisfying $\Delta F = \pm2$, referred to as magnetically-induced transitions, can be observed. Here, we show that
%at intermediate magnetic field,
%$B\sim B_0 = A_\text{hfs}/\mu_B$, $A_\text{hfs}$ where $A_\text{hfs}$ is the magnetic dipole constant, and $\mu_B$ is the Bohr magneton ($B_0\approx 1.7$\,kG for Cs) 
a significant redistribution of the Cs $6\text{S}_{1/2}\rightarrow 6\text{P}_{3/2}$ magnetically-induced transition intensities occurs with increasing magnetic field. We observe that the strongest transition in the group $F_g=3\rightarrow F_e=5$ ($\sigma^+$ polarization) for $B<B_0$ cease to be the strongest for $B>3\,B_0$. On the other hand, the strongest transition in the group $F_g=2\rightarrow F_e=4$ ($\sigma^-$ polarization) remains so for all our measurements with magnetic fields up to 9\,kG. These results are in agreement with a theoretical model.
%based on the diagonalization of the Zeeman Hamiltonian matrix
The model predicts that similar observations can be made for all alkali metals, including Na, K and Rb atoms.
Our findings are important for magnetometers utilizing the Zeeman effect above Earth field, following the rapid development of micro-machined vapor-cell-based sensors. 
\end{abstract}


\end{frontmatter}

%\linenumbers

\section{Introduction}

Hot atomic vapors of alkali metal atoms are at the core of many fundamental and applied precision physics experiments. They can be used in the search for signatures of beyond-the-Standard-Model physics \cite{safranoveRMP2018}, and possess a large range of applications.
%, for example biomagnetism \cite{knappe2014optically}, positioning systems \cite{vig1993military}.
In contrast to other technologies with similar performances, vapor-cell-based sensors have been receiving a lot of attention because of the possibilities for miniaturization and low power consumption while retaining their sensitivities \cite{kitching2018chip}.

% Puisque beaucoup de phénomènes physiques influencent la structure atomique donnant lieu aux signaux spectroscopiques, et puisque ces signaux peuvent être mesurés très précisément, la réalisation de capteurs à hautes performances est possible à l'aide de cellule de vapeurs \cite{kitching2011Atomic}. Ceux-ci rivalisent même avec d'autres types de capteurs (atomes froids, piège à ions, etc.) qui sont eux bien plus compliqués, dépendants de cryogènes, énergivores et/ou volumineux. For example, optically-pumped magnetometers with a sensitivity have been demonstrated \cite{kominis2003subfemtotesla}, as well as nuclear magnetic resonance gyroscopes \cite{kornack2005Nuclear} and stabilized frequency references \cite{newman2021high}.  \cite{kitching2018chip}.

The Zeeman effect occurs when atoms are in the presence of an external magnetic field. Hyperfine levels are split, and transition frequencies and intensities undergo significant changes as a function of the field magnitude \cite{tremblayPRA1990,auzinshBook2010}. At intermediate magnetic field amplitude $B\sim B_0 = A_\text{hfs}/\mu_B$, where $A_\text{hfs}$ is the magnetic dipole constant of the ground state, and $\mu_B$ is the Bohr magneton ($B_0\approx 1.7$\,kG for Cs), the coupling between magnetic sub-levels induced by the field leads to an ambiguous set of transition selection rules. Transitions satisfying $\Delta F = \pm2$, referred to as magnetically-induced transitions (MI), can be observed while being forbidden at zero field. As demonstrated in previous studies \cite{tonoyanEPL2018,sargsyanPhysLetA2021}, the intensity of MI transitions satisfying the condition $F_e- F_g \equiv \Delta F = +2$ is the greatest with $\sigma^+$ polarized radiation, while MI transitions satisfying the condition $ \Delta F = -2$ exhibit the greatest intensity with $\sigma^-$ polarized radiation. 
% Since these transitions are dipole-forbidden at $B = 0$, there are no radiation absorption or fluorescence processes at these transitions. However, in a magnetic field, a gigantic increase occurs for both absorption and fluorescence, and they are therefore called magnetically induced (MI) transitions.

At low and intermediate field, these transitions are often overlooked, as conventional weak-probe spectroscopy is not enough to properly resolve individual transitions because of the Doppler effect \cite{pizzey2022laser}. On the other hand, sub-Doppler spectra obtained with non-linear techniques can be difficult to interpret. With nanocells (NCs), however, one can directly obtain sub-Doppler spectra while remaining in the weak-probe regime of interaction. In these cells, the vapor is confined between two almost-parallel windows with a separation spanning from a few nm to a few $\mu$m. The confinement of the vapor to dimensions on the order of the probe optical wavelength leads to the (optical) Dicke-type coherent narrowing \cite{dutier2003collapse}, well known in the microwave domain \cite{romer1955new}. In addition to spectroscopy, NC are an important tool in fundamental physics: they can for example be used to study atom-surface interactions \cite{Laliotis2021atom-surface,sargsyanPLA2023}, cooperative effects \cite{keaveney2012cooperative,peyrot2019Optical}, and single photon generation \cite{ripka2018room}.

Magnetically-induced transitions can be utilized as novel frequency markers in high magnetic fields, enabling the exploration of new frequency ranges and facilitating the frequency stabilization of lasers operating at significantly shifted frequencies compared to the initial transitions in unperturbed atoms \cite{sargsyanOL2014}. Magnetically-induced transitions in Cs and Rb have also found successful applications in the process of electromagnetically induced transparency in strong magnetic fields \cite{sargsyan2019dark,sargsyan2023formation,sargsyan2023application}. The complete understanding of the evolution of alkali transitions in a magnetic field is important for high field magnetometry \cite{GeorgeRSI2017,ciampini2017optical,keaveney2019quantitative,klingerAO2020,staerkind2023high}. As these magnetometers are based on optical readout, they are compatible with harsh environments applications \cite{FuAQS2020}. Consequently, it is important to investigate the behavior of MI transitions
%$3 \rightarrow 5'$ and $4 \rightarrow 2'$ 
in magnetic fields exceeding $B_0$. 

In this article, we study the absorption spectrum of Cs atoms confined in a NC, and placed in a magnetic field spanning over a few kG. We show that the strongest transition in the group $F=3\rightarrow 5'$ ($\sigma^+$ polarization) for $B<B_0$ cease to be the strongest for $B>3\,B_0$. On the other hand, the strongest transition in the group $F=2\rightarrow 4'$ ($\sigma^-$ polarization) remains so for all magnetic field values, in agreement with a theoretical modeling of thin vapor layer absorption spectrum of alkali atoms placed in a magnetic field, see Sec.\,\ref{sec:theory}. The experimental setup and results are presented in Sec.\,\ref{sec:experiment}.

\section{Spectrum of Cs atoms in a magnetic field}\label{sec:theory}

\subsection{Absorption profile}

The absorption profile of atoms confined in nanocells is fundamentally different from that of a usual vapor cell. In the case of a nanocell,
%with a low optical density (typ. $T< 200\,^\circ$C)
 the atomic response is dominated by a
%a homodyne beating between the Fabry-Pérot signal (because of parallel windows) and that of light scattered by the atoms
large spatial dispersion \cite{ermolaevPRA2022}. The profile strongly depends on the thickness $\ell$ of the probed vapor layer \cite{dutier2003collapse}. In the weak-probe regime, one can model the system as an ensemble of two-level systems interacting with light of angular frequency $\omega_l$ and wavevector $k$, see e.g. Ref.\,\cite{sargsyanPhysLetA2021}. Each two-level system has a transition frequency
%$\omega_a$ 
and a dipole moment 
%$\langle e||d||g\rangle$ 
that depend on the magnitude of the magnetic field. The transmission spectrum is then obtained by summing over all possible transitions $j$:
\begin{equation}
    S_t \propto \frac{1}{|Q|^2}\cdot\sum_j\text{Im}\left[\chi_j(\omega_l,\ell,B)\right] ,
\end{equation}
where $Q=1-r^2\exp(2ik\ell)$ and $r$ is the (field) reflection coefficient of the cell windows. An analytical expression for the transition lineshape of a two-level system was derived in Ref.\,\cite{dutierJOSAB2003}, it reads
\begin{equation}
    \chi_j(\omega_l,\ell,B)=-4(1-r\exp[ik\ell])^2\cdot\frac{\sin(k\ell/2)}{Q}\cdot\frac{\mathcal{N}}{ku\sqrt{\pi}}\cdot\frac{iA_j}{\Gamma/2-i\Delta_j},
\end{equation}
where $\mathcal{N}$ is the vapor density, $u(\Theta)=\sqrt{2k_B\Theta/m_a}$ is the thermal velocity at a temperature $\Theta$ for atoms of mass $m_a$. The (magnetic-field dependent) transition parameters $\Delta_j=\omega_l-\omega_j$  and $A_j$ (proportional to the squared dipole moment) are derived in the next section. The homogenous broadening $\Gamma$, including contributions from natural linewidth, collisional broadening, etc., is left as a free parameter in our simulations.

\subsection{Zeeman effect}
The theoretical description of the Zeeman effect on the spectrum of alkali atoms is extensively detailed in various studies \cite{tremblayPRA1990,alexandrov1993interference,auzinshBook2010,sargsyanLPL2014,scottoPRA2015,sargsyanJOSAB2017}. In the presence of magnetic field, the magnetic sublevels are split leading to changes in transition probabilities and frequencies. This can be calculated by diagonalizing the Hamiltonian matrix accounting for the hyperfine atomic structure, and the interaction with the magnetic field
\begin{equation}
    H = H_{\text{hfs}} +\frac{\mu_B}{\hbar}(g_S\,S_z +g_L\,L_z + g_I\,I_z) B_z\,,
\end{equation}
where $g_{S,L,I}$ are the Landé factors \cite{olsen2011optical,staerkind2023precision} and $S_z$, $L_z$, $I_z$ the projection of quantum numbers on the $z$-axis (quantization axis).
The eigenvalues give the energy levels, needed to calculate the transition frequencies, and the eigenvectors are used to calculate the transition dipole moments
\begin{equation}
    |\langle e||d||g\rangle|^2 \propto \Gamma_\text{N} \, a^2[\psi(F_e',m_{F_e});\psi(F_g',m_{F_g});q]\,,
\end{equation}
with $\Gamma_\text{N}$ the natural linewidth of the transition. The transfer coefficients are calculated with
\begin{equation}
a[\psi(F_e',m_{F_e});\psi(F_g',m_{F_g});q] = \sum_
{F_e,F_g}C_{F'_eF_e}a(F_e',m_{F_e};F_g',m_{F_g};q)C_{F'_gF_g}\,,
\end{equation}
 where $C_{F'F}$ are the mixing coefficients (obtained from the eigenvectors). The coefficient $a(F_e',m_{F_e};F_g',m_{F_g};q)$ reads 
\begin{equation}
\begin{aligned}
a(F_e',m_{F_e};F_g',m_{F_g};q)=(-1)^{1+I+J_e+F_e+F_g-m_{Fe}}\sqrt{2J_e+1}\sqrt{2F_e+1}\\ \times \sqrt{2F_g+1} \left( \begin{array}{r@{\quad}cr} 
F_e & 1 & F_g \\
-m_{F_e} & q & m_{F_g}
\end{array}\right)
\left\{ \begin{array}{r@{\quad}cr} 
F_e & 1 & F_g \\
J_g & I & J_e
\end{array}\right\},
\end{aligned}
\end{equation}
where $q=0,\pm1$ is associated with the polarization of the incident laser field. In the previous equation, the parentheses (curly brackets) denotes 3-$j$ (6-$j$) coefficients.

\begin{figure}[htb]
    \centering
    \includegraphics[scale=1]{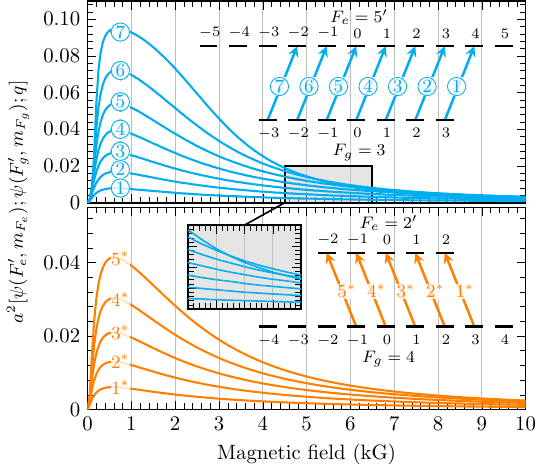}
    \caption{Calculated probabilities of $F_g=4\rightarrow 2'$ transitions ($q=-1$, bottom) and $F_g=3\rightarrow 5'$ transitions ($q=1$, top) plotted against magnetic field. The inset plot shows a zoom on magnetic fields ranging from 4.5 to 6.5\,kG. Transition diagrams and labeling are shown as insets.}
    \label{fig:MI-theory}
\end{figure}

\begin{figure}[htb]
    \centering
    \includegraphics[scale=1]{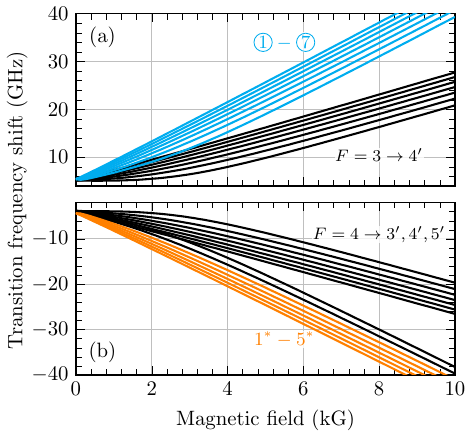}
    \caption{Transition frequency shift as a function of the magnetic field. (a) $F=3 \rightarrow 4',5'$ transitions in the case of $\sigma^+$-polarized light, including the \textcircled{1} -- \textcircled{7} MI transitions (cyan). (b) $F=4 \rightarrow 2',3',4',5'$ transitions in the case of $\sigma^-$-polarized light, including the $1^*$ -- $5^*$ MI transitions (orange). The frequency shifts are calculated with respected to the weighed center of the Cs D$_2$ line.}
    \label{fig:MI-shift}
\end{figure}

The calculated MI transition intensities are shown in Fig.\,\ref{fig:MI-theory}. A striking example of a gigantic increase in transition intensity is the behavior of the $F= 3 \rightarrow 5'$ transitions (seven transitions) and $F=2 \rightarrow 4'$ (five transitions) of the Cs D$_2$ line. Note that the excited state quantum numbers are denoted with primes. In the range of $0.5-1$\,kG, the intensities of these transitions increase greatly, especially for $\sigma^+$ circularly polarized radiation. Three of them (\textcircled{5},\textcircled{6}, and \textcircled{7}) have the highest probabilities in the range of $0.5-2$\,kG among all transitions originating from $F_g=3$ \cite{sargsyanLPL2014}. Magnetically-induced transitions have interesting characteristics: 
$(i)$ they remain observable with magnetic fields as high as $B_z \sim 8$\,kG;
$(ii)$ they are located either in the high-frequency wing of the spectrum (for $F=3\rightarrow 5'$) or in the low-frequency wing of the spectrum (for $F=4 \rightarrow 2'$), see Fig.\,\ref{fig:MI-shift}, without intersecting much with other Cs transitions;
$(iii)$ for $B_z \sim 8$\,kG, the frequency shift of transitions reaches about $\pm 35$\,GHz in the case of $\sigma^\pm$ polarization,  with respect to the weighed center of the Cs D$_2$ line.

% A similar behavior is observed for magnetically induced transitions from $F_g=4 \rightarrow F_e=2$ (five transitions) of Cs D$_2$ line atoms in an external magnetic field, particularly for $\sigma^-$ circularly polarized radiation, as indicated by numbers $1^*-5^*$ inside rectangles in Figure\,\ref{fig:MI-theory}.

 From Figure\,\ref{fig:MI-theory}, we can see that transition $5^*$ remains the strongest one in the $4\rightarrow 2'$ group, in the whole magnetic field range where it is measurable. On the other hand, transition \textcircled{7} is the strongest one only up to about $3B_0 \approx 5.1\,$kG, see the inset in Fig\,\ref{fig:MI-theory}. Above $5B_0$, two MI transitions emerge as the strongest in the group of $F=3\rightarrow5'$: transitions \textcircled{6} and \textcircled{5}. 

The magnetic field dependence of the frequency shift for the seven MI transitions $F=3\rightarrow 5'$ and five transitions $F=4\rightarrow 2'$ is depicted in Figure\,\ref{fig:MI-shift}. Beyond $B_z=3.5$\,kG, both groups of transitions become distinctly separated, which holds significance for applications involving MI transitions, for example for laser frequency stabilization \cite{sargsyanOL2014}.

\section{Experiment} \label{sec:experiment}

\begin{figure}[tb]
    \centering
    \includegraphics[width=0.8\textwidth]{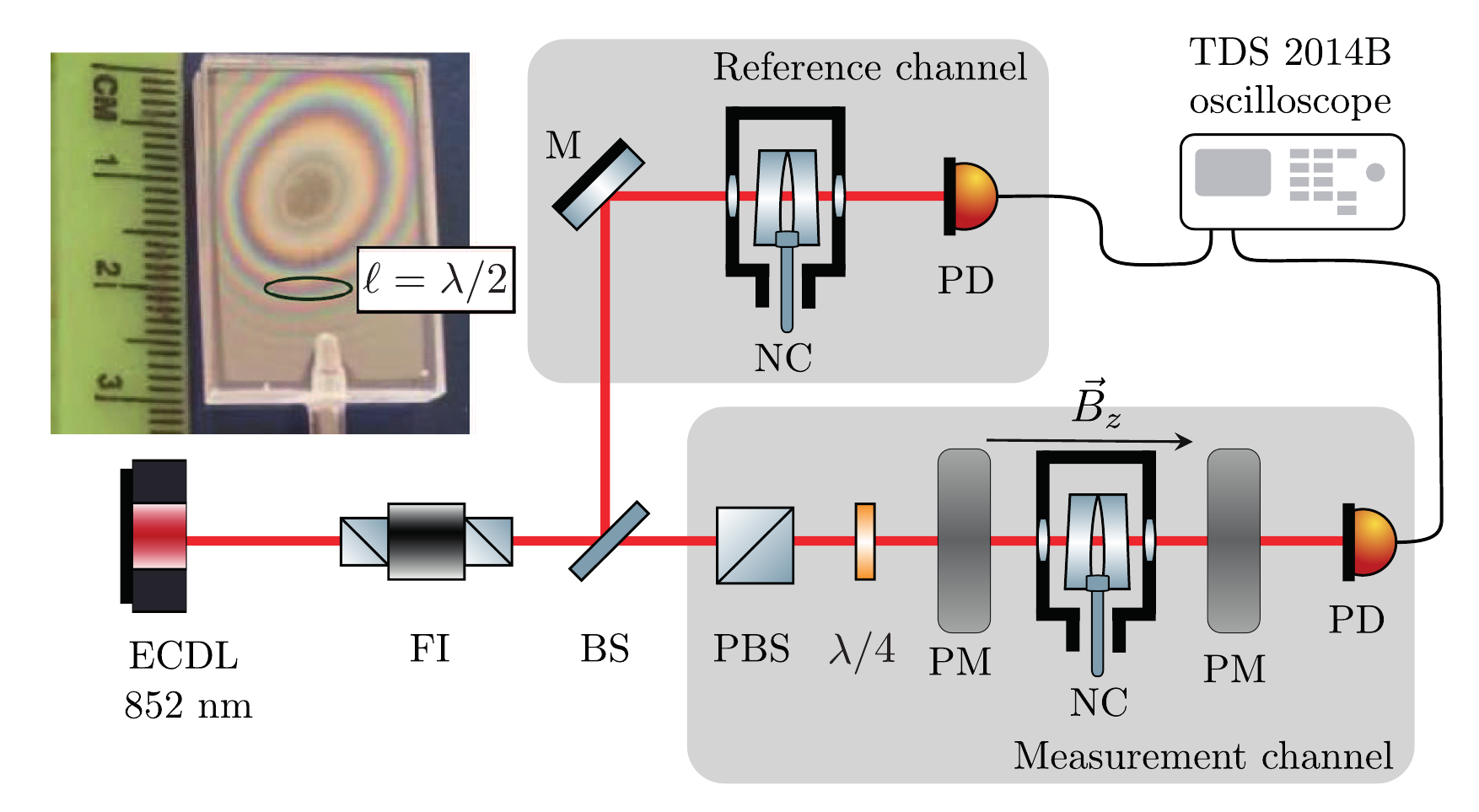}
    \caption{Sketch of the experimental setup. ECDL -- extended cavity diode laser, FI -- Faraday isolator, BS -- beam splitter, M -- mirror, PBS -- polarizing beam splitter, NC -- Cs nanocell in the oven, $\lambda/4$ -- quarter-wave plate, PM -- permanent ring magnets, PD -- photodetectors.}
    \label{fig:setup}
\end{figure}

The experimental setup is depicted in Figure\,\ref{fig:setup}. A tunable external-cavity diode laser (ECDL) emitting at a wavelength $\lambda=852$\,nm with a spectral linewidth of less than 1\,MHz is tuned in the vicinity of the Cs D$_2$ line. Single-beam sub-Doppler spectroscopy is achieved by illuminating  the nanocell with a 1\,mm-diameter laser beam in the region having a thickness $\ell=\lambda/2\approx 426\,$nm \cite{dutier2003collapse,sargsyanOL2019}. This region is indicated in the inset of Figure\,\ref{fig:setup}. In addition to increasing the spectral resolution, using a NC also allows to record the spectrum in the linear regime of interaction between light and atoms. This is convenient to directly compare the amplitude of the absorption peaks and extract ratios of amplitudes. The cell temperature is set to $100^\circ\,$C, and the power of the laser beam was about $15\,\mu$W.
%atomic media in transmission experiments, and this was accomplished using an optical vapor nanocell (NC). The NC possessed a vapor column thickness equal to half of the wavelength of the resonant laser radiation , resulting in a vapor column thickness $\ell \sim \lambda/2 = 426$\,nm.
% Further design details of the NC can for example be found in Ref.\,\cite{KeaveneyPRL2013}.

Magnetic fields ranging from $B_z = 1$\,mT to 0.9\,T were applied using a robust permanent ring magnets system mounted on a translation stage \cite{sargsyanJETP2015}. Despite a large gradient (with respect to e.g. a pair Helmholtz coils), the spectral resolution is acceptable when using permanent magnets: transitions are additionally broadened by only a few tens of MHz. This is because of the small thickness of probed layer 
\cite{sargsyanApplPhysLet2008}.
The magnetic field was aligned with the propagation direction of the laser radiation, so that $\sigma^\pm$ transition can be observed. The light polarization was adjusted using a quarter-wave plate placed before the NC-oven assembly. Approximately 30\% of the laser radiation was split off to form a frequency reference channel. The change of light intensity by propagating through the nanolayer of alkali was captured with photodiodes whose signal was fed to a four-channel oscilloscope (Tektronix TDS2014B).

%\subsection{Magnetically-induced transitions $F_g=3 \rightarrow F_e=5$}

\begin{figure}[htb]
    \centering
    \includegraphics[scale=1]{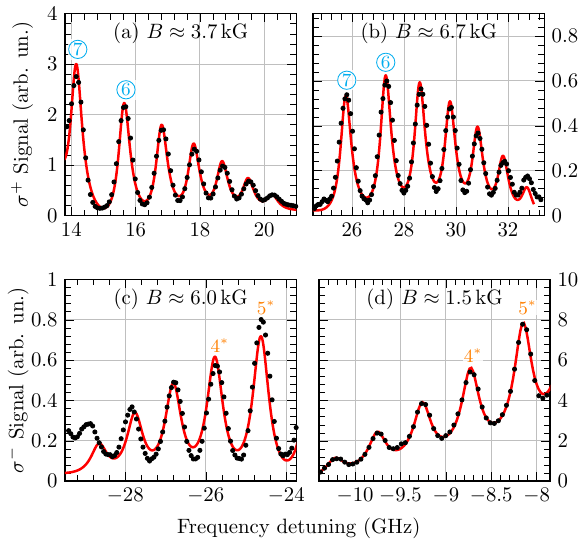}
    \caption{Examples of Cs D$_2$ line absorption spectrum in a magnetic field for (a-b) $\sigma^+$-polarized light, (c-d) $\sigma^-$-polarized light recorded with 32 averages. The model discussed in Sec.\,\ref{sec:theory} (red lines) was fit to the experimental data (black dots). Depending on the magnitude of the magnetic field, fitted linewidths ranging from $50-75\,$MHz were extracted from the fits. The zero frequency detuning corresponds to the frequency of the that of the 6S$_{1/2}\rightarrow6$P$_{3/2}$ transition.}
    \label{fig:Exp-sigmaMinus}
\end{figure}

Figure\,\ref{fig:Exp-sigmaMinus} shows examples of absorption spectra recorded from the nanocell at various magnitude of the magnetic field for $\sigma^+$ (top) and $\sigma^-$ (bottom) laser polarization. In these spectra, all 12 MI transitions (\textcircled{1} to \textcircled{7} and $1^*$ to $5^*$) for both circular polarizations are observed and well resolved. Fits of the model described in Sec.\,\ref{sec:theory} to experimental data (black dots) are shown with red solid lines. A good agreement between experiment and theory can be seen.
% These spectra correspond to MI transitions from 3 to 5', with $\sigma^+$ polarization, observed in strong magnetic fields of 3.7 kG (left side) and 6.7 kG (right side).
%These measurements were conducted using the nanocell with a thickness $\ell \sim \lambda/2 = 426$\,nm.
For $B_z < 3 B_0$ [Fig.\,\ref{fig:Exp-sigmaMinus}(a)], the transition $|3,-3\rangle \rightarrow |5',-2'\rangle$, labeled \textcircled{7}, is seen to be larger than $|3,-2\rangle \rightarrow |5',-1'\rangle$, labeled \textcircled{6}. Note that states are expressed in the coupled basis of states $|F,m_F\rangle$; for the complete set of labels, see the inset in Fig.\,\ref{fig:MI-theory}. However, when $B_z > 3 B_0$ [Fig.\,\ref{fig:Exp-sigmaMinus}(b)], transition \textcircled{6} is seen to be the largest one among the seven $\sigma^+$ MI transitions of this group. In contrast, the transition $|4,-1\rangle \rightarrow |2',-2'\rangle$, labeled $5^*$ remains the strongest one, [see Fig.\,\ref{fig:Exp-sigmaMinus}(c) and (d)].

% The most prominent MI transition corresponds to the lowest value of $m_F$ for the $F_g$ level \cite{sargsyanJPB2020}, which is $m_F = -3$, corresponding to transition number 7 indicated within a circle. Using a representation like $|F, m_F\rangle$ (coupled basis), transition number 7 in the circle can be expressed as $|3,-3\rangle \rightarrow |5',-2'\rangle$. This description holds true for magnetic fields up to , where $B_0 (Cs) = 1.7 $\,kG.

\begin{figure}
    \centering
    \includegraphics[scale=1]{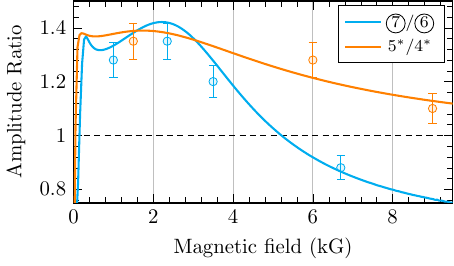}
    \caption{Evolution of transition amplitude ratio as a function of applied magnetic field for the two strongest $\sigma^+$ (cyan dots) and $\sigma^-$ (orange dots) MI transitions. The amplitude of the error bars corresponds to the measured voltage noise on the peaks of interest.  The lines show the theoretical ratio. The case where the amplitude of both transitions are equal is indicated with a dashed line.}
    \label{fig:5}
\end{figure}

The magnetic-field dependence of the ratio between the strongest MI transitions in the case of $\sigma^+$- (cyan) and $\sigma^-$- (orange) polarized light is shown in Fig.\,\ref{fig:5}. The experimental data (dots) were obtained by simply dividing the on-resonance signal of the transitions. The amplitude of the error bars corresponds to the measured voltage noise on the peaks of interest. Below $3B_0\approx 5.1\,$kG, transitions \textcircled{7} and $5^*$ have the largest amplitude. However, one can observe that, above $3B_0$, transition \textcircled{6} becomes larger than \textcircled{7} (ratio below 1). Transition $5^*$ remains stronger than $4^*$ in the whole range of measurements. Despite the simple method used to extract the peak amplitude, these observations align with the calculated ratios of MI transition intensities as a function of magnetic field (solid lines in Fig.\,\ref{fig:5}). 
This can be useful, for example for large field magnetometers utilizing quick algorithms to find the peak heights and positions and deduce the magnetic field seen by the atoms.

% The experimental (the upper curve) and theoretical (the lower curve) absorption spectra for MI transitions $4 \rightarrow 2'$ , $\sigma^-$ polarization in strong magnetic field 6 kG and 1.5 kG, shown in the left and right sides respectively, are shown in Fig.8. The amplitude of the MI transition $|4,-1\rangle \rightarrow |2',-2'\rangle$ with number 5 in a rectangle is the largest for all B values. This is in agreement with the theoretical curves shown in Fig. 2. This is consistent also with the experimental and theoretical ratio of the absorption amplitudes $A(5)$ to $A(4)$ versus B for the $\sigma^-$ excitation shown in Fig. 6.

\section{Conclusion}

 We have studied the behavior of magnetically-induced transitions in magnetic fields up to 9\,kG by means of linear transmission spectroscopy from a nanocell. A large redistribution of the intensities with increasing magnetic field  is observed. In particular, the $\sigma^+$ MI transition $|3,-3\rangle \rightarrow |5',-2'\rangle$ cease to be the most intense at field above $3\,B_0$. In contrast, the $\sigma^-$ MI transition $|4,-1\rangle \rightarrow |2',-2'\rangle$ is observed to remain the strongest in the whole range of investigated magnetic field. A simple reading of the on-resonance signal is found to yield results in agreement with a theoretical modeling of thin vapor layer absorption spectrum of alkali atoms placed in a magnetic field. The model predicts a similar behavior for MI transitions of other alkali metals, including Na, K and Rb atoms. For example, the $^{85}$Rb $|2,-2\rangle\rightarrow |4',-1'\rangle$ MI transition cease to be the strongest one in the group $F=2\rightarrow 4'$ above $3B_0(^{85}\text{Rb})\sim 2.1\,$kG while the MI transition $|3,0\rangle\rightarrow |1',-1'\rangle$ remains the strongest one in the group $F=3\rightarrow 1'$.

These results could be beneficial to large field magnetometers utilizing the Zeeman effect. With quick algorithms to find the peak heights and positions to deduce the magnetic field seen by the atoms, one could improve the bandwidth of high-dynamic range magnetometers based on alkali D line spectrum fitting \cite{klingerAO2020}. Because of a smaller sensitivity to field gradient, one could achieve better sensitivities in high pulsed field measurements \cite{GeorgeRSI2017} by using nanocells.
To conclude, we note that our findings could also be useful to enlarge the tuning range of group velocity manipulation for slow-light experiments using electromagnetically-induced transparency \cite{NguyenOptik2018}.  
	
\section*{Acknowledgement}
The authors would like to thank A. Amiryan for fruitful discussions. The work was supported by the Science Committee of RA, in the frame of the research project No 21T-1C005. 

\section*{References}
\bibliography{Bib-main}

\end{document}